\documentstyle[psfig,draft]{mn}

\def\gs{\mathrel{\raise0.35ex\hbox{$\scriptstyle >$}\kern-0.6em
\lower0.40ex\hbox{{$\scriptstyle \sim$}}}}
\def\ls{\mathrel{\raise0.35ex\hbox{$\scriptstyle <$}\kern-0.6em
\lower0.40ex\hbox{{$\scriptstyle \sim$}}}}

\def\h1{{\rm h}^{-1}}

\title[A Multiply-imaged ERO at $z=1.60$]
{A \textit{Hubble Space Telescope} Lensing Survey of 
X-ray Luminous Galaxy Clusters:  III. \newline
A Multiply-imaged Extremely Red Galaxy at \textit{z}=1.6}

\author[Smith et al.]
{Graham P.\ Smith,$^{\! 1}$\footnotemark[1] Ian Smail,$^{\! 1}$ J.-P.
Kneib,$^{\! 2}$ C.J.\ Davis,$^{\! 3}$ M.\ Takamiya,$^{\! 4}$ \and 
H.\ Ebeling~$^{\! 5}$ \& O.\ Czoske~$^{\! 2}$
\vspace*{1mm}\\
$^1$ Department of Physics, University of Durham, South Road, Durham DH1 3LE\\
$^2$ Observatoire Midi-Pyr\'en\'ees, 14 Avenue E.\ Belin, 31400 Toulouse, 
France\\
$^3$ Joint Astronomy Centre, 660 North A'ohoku Place, University Park, Hilo, 
HI 96720, USA\\
$^4$ Gemini Observatory, 670 North A'ohoku Place, University Park, Hilo, HI 96720, USA\\
$^5$ Institute for Astronomy, University of Hawaii, 2680 Woodlawn Drive, 
Honolulu HI\,96822, USA\\
}

\pagerange{000--000}

\begin{document}

\maketitle

\begin{abstract}

We present near-infrared spectroscopy  and {\it Hubble Space Telescope
(HST)}  imaging  of  ERO\,J003707$+$0909.5,  the  brightest  of  three
gravitationally-lensed  images of  an  Extremely Red  Object (ERO)  at
$z=1.6$, in  the field of  the massive cluster A\,68  ($z=0.255$).  We
exploit  the superlative  resolution of  our \emph{HST}  data  and the
enhanced  spatial  resolution and  sensitivity  afforded  by the  lens
amplification to reconstruct the  source-plane properties of this ERO.
Our morphological  and photometric analysis  reveals that ERO\,J003707
is an $L^{\star}$ early-type disk-galaxy and we estimate that $\sim10$
per  cent of  EROs with  $(R-K)\ge5.3$ and  $K\le21$ may  have similar
properties.  The  unique association  of passive EROs  with elliptical
galaxies therefore appears to be  too simplistic.  We speculate on the
evolution of ERO\,J003707:  if gas continues to cool  onto this galaxy
in the manner predicted  by hierarchical galaxy formation models, then
by the  present day,  ERO\,J003707 could evolve  into a  very luminous
spiral galaxy.

\end{abstract}


\section{Introduction}

\footnotetext[1]{E-mail: graham.smith@durham.ac.uk}

The  cores of massive  galaxy clusters  act as  powerful gravitational
lenses, providing  a magnified view of  serendipitously placed distant
galaxies.   Indeed, close alignment  of a  background galaxy  behind a
cluster  lens  can  lead  to  multiple  images  being  visible.   This
phenomenon,  known as  strong lensing,  has been  applied  on numerous
occasions  to the  study of  high-redshift galaxies  (e.g.\  Hammer \&
Rigaut 1989 ;  Ebbels et al.\ 1996; Franx et al.\  1997; Seitz et al.\
1998; Pell\'o et al.\ 1999; Ellis  et al.\ 2001; Smith et al.\ 2001a).
A key feature of these  studies is that the lens magnification enables
intrinsically low-luminosity galaxies at  remote epochs to be observed
and studied in great detail.

One  class  of distant  galaxy  that  may  benefit from  the  enhanced
sensitivity provided by a  gravitational lens is Extremely Red Objects
(EROs).   The   defining   optical/near-infrared  colours   of   EROs,
$(R-K)\ge5.3$, should select either distant elliptical galaxies (e.g.\
Dunlop et  al.\ 1996;  Soifer et al.\  1999) or  heavily dust-obscured
systems at $z\sim1$--2 (Dey et  al.\ 1999; Pierre et al.\ 2001; Afonso
et al.\ 2001; Smith et al.\ 2001b).  Here we concentrate on those EROs
whose $(R-K)$  colour is dominated  by the $4000{\rm\AA}$ break  of an
evolved stellar population  (i.e.\ ``passive EROs'').  Various authors
(e.g.\ Kauffmann \& Charlot 1998;  Fontana et al.\ 1999; Daddi et al.\
2000) have used  observations of passive EROs  and other $K$--selected
samples to  argue for or  against monolithic collapse (e.g.\  Eggen et
al.\ 1962; Larson 1975; Tinsley \& Gunn 1976) or hierarchical assembly
(e.g.\  White \&  Frenk 1991;  Cole et  al. 2000)  theories  of galaxy
formation.  Regardless  of the details of this  debate, both arguments
implicitly  assume   that  \emph{all}  passive   EROs  are  elliptical
galaxies.

The importance  of EROs to  our understanding of galaxy  formation and
the intrinsic faintness of these systems ($R\gs23$, $K\gs18$) recently
motivated us to search for EROs in the fields of massive gravitational
lenses  (Smith et al.\  2002a --  S02a).  In  this letter,  we present
near-infrared spectroscopy of ERO\,J003707$+$0909.5, drawn from S02a's
sample.   We  describe  our  observations  in \S2,  present  our  lens
modelling in  \S3, describe our source-plane analysis  in \S4, discuss
the diversity  of passive EROs in  \S5, speculate on  the evolution of
ERO\,J003707 in \S6 and finally  summarize our conclusions in \S7.  We
assume $H_0=50$kms$^{-1}$Mpc$^{-1}$, $\Omega_0$=1 and $\Lambda_0=0$.

\section{Observations}

\setcounter{footnote}{1}

ERO\,J003707  was  first  detected  in \emph{HST}  \footnote{Based  on
observations  with the NASA/ESA  Hubble Space  Telescope at  the Space
Telescope Science  Institute, which is operated by  the Association of
Universities for Research in  Astronomy, Inc., under NASA contract NAS
5-26555.}  imaging  of A\,68  as part of  our lensing survey  of X-ray
luminous galaxy  clusters (Smith  et al.\ 2001a;  S02a; Smith  et al.\
2002b,   in   preparation  --   S02b).    A\,68   was  observed   with
\emph{HST}/WFPC2  for 7.5\,ks  through  the F702W  filter and  8.8\,ks
using UFTI on the 3.8-m United Kingdom Infrared Telescope\footnote{The
United Kingdom  Infrared Telescope is operated by  the Joint Astronomy
Centre  on  behalf of  the  Particle  Physics  and Astronomy  Research
Council}  (UKIRT).  We  show  a subset  of  these data  in Fig.~1  and
identify the  three images  of ERO\,J003707.  We  refer the  reader to
S02a and S02b for further details of these data and their reduction.

Our  near-infrared spectroscopy  targeted  specific spectral  features
based on  the likely nature of  ERO\,J003707, and a  crude estimate of
its  redshift.  We  first examined  S02a's $RIJK$--band  photometry of
ERO\,J003707 and found that it does not discriminate between old stars
and  dust-obscuration as  the origin  of the  extreme colours  of this
galaxy.   For simplicity,  we  therefore fitted  an elliptical  galaxy
spectral template (Coleman  et al.\ 1980) to S02a's  photometry of the
central red  component using {\sc  hyper-z} (Bolzonella et  al.\ 1999)
and obtained  $z_{\rm phot}\sim1.7$.  Assuming  that the near-infrared
emission is dominated by old stars, we searched for the $4000{\rm\AA}$
break and  CaII H and K  absorption lines in  the $z$--band.  However,
since S02a's photometry is not conclusive, and ERO\,J003707 appears to
be a disk-galaxy (\S4), we  also observed in the $J$-- and $H$--bands,
searching for H$\beta$, [OIII],  H$\alpha$ and [NII] emission lines in
addition to [OII] which should appear in the $z$--band if ERO\,J003707
is a star-forming galaxy at $z\sim1.7$.

We observed ERO\,J003707 with CGS4 (Mountain et al.\ 1990) on UKIRT on
2001  September  15--18 in  non-photometric  conditions and  $\sim1''$
seeing  with  the  slit   centered  on  the  $K$--band  emission  from
ERO\,J003707$+$0909.5                            ($\alpha,\delta\,({\rm
J2000})=00\,37\,07.37+09\,09\,28.4$).   We  adopted  a  slit  position
angle of $22^{\circ}$ to ensure that the $K$--band emission from image
B (Fig.~1)  also fell on the  slit, although this meant  that the disk
light was not well sampled.  Nodding along the slit, we obtained total
integrations  of 7.2\,ks  and  21.6\,ks in  the  $J$-- and  $H$--bands
respectively.  These data were reduced in a standard manner using {\sc
iraf} tasks.   The final reduced  frames contained no  strong spectral
features, although a weak continuum was detected in the $H$--band.  We
also used  the Keck-II 10--m  telescope\footnote{Based on observations
made at the  W. M. Keck Observatory by Gemini  staff, supported by the
Gemini  Observatory,   which  is   operated  by  the   Association  of
Universities  for  Research  in  Astronomy,  Inc., on  behalf  of  the
international  Gemini  parthership.   The  W.M.  Keck  Observatory  is
operated as a scientific partnership among the California Institute of
Technology, the University of  California and the National Aeronautics
and Space  Administration.} to obtain  $z$-- and $H$--band  spectra of
ERO\,J003707$+$0909.5  with  NIRSPEC (McLean  et  al.,  1998) on  2001
November 20--21.  Using the same target co-ordinates and slit position
angle,  we  integrated  for  7.2\,ks  and 3.6\,ks  in  the  $z$--  and
$H$--bands   respectively,  in   clear   and  photometric   conditions
(FWHM$\sim0.8''$).  These data were  also reduced in a standard manner
using {\sc iraf} tasks,  including the {\sc wmkonspec} package.  These
data   confirm  the   absence  of   strong  emission   lines   in  the
$H$--band. The  $z$--band data also contain no  strong emission lines,
however they  reveal a spectral  break at $1.04\pm0.01\mu$m.   We show
the  $z$--band  continuum in  Fig.~2.   Given  the  absence of  strong
emission lines in the $zJH$--bands  and the fact that our observations
primarily  sample  the  central  red  component  of  this  galaxy,  we
interpret the break  in the $z$--band spectrum of  ERO\,J003707 as the
$4000{\rm\AA}$ break at a redshift of $z=1.60\pm0.03$.

\section{Gravitational Lens Modelling}

The detailed construction  of the lens model of  A\,68 is described by
S02b; here we summarize its key features.  The model contains a single
lens  plane at  the cluster  redshift ($z=0.255$)  comprising  36 mass
components:  the cluster  dark  matter halo  centered  on the  central
galaxy; a second dark matter  halo centered on the brightest galaxy of
a  group lying  $\sim80''$ North-West  of the  cluster center;  and 34
cluster  galaxies including  the  central galaxy.   Each component  is
described by the  parameters: position, ellipticity, orientation, core
radius ($r_{\rm  core}$), cut-off  radius ($r_{\rm cut}$)  and central
velocity  dispersion ($\sigma_{\rm  o}$).   The analytical  expression
used   to   describe   each   component  is   a   smoothly   truncated
pseudo-isothermal elliptical mass distribution  (Kneib et al.\ 1996 --
K96).  The center, ellipticity  and orientation of each mass component
is matched to  the observed light distribution of  the related cluster
elliptical.  The  dynamical parameters ($r_{\rm  core}$, $r_{\rm cut}$
and $\sigma_{\rm  o}$) of  the main cluster  dark matter halo  and the
central  velocity dispersion  of  the  N-W dark  matter  halo and  the
central galaxy  are kept as free parameters.   The remaining dynamical
parameters are  scaled with the luminosity of  their associated galaxy
following K96.

%
%
\begin{figure}
\centerline{\psfig{file=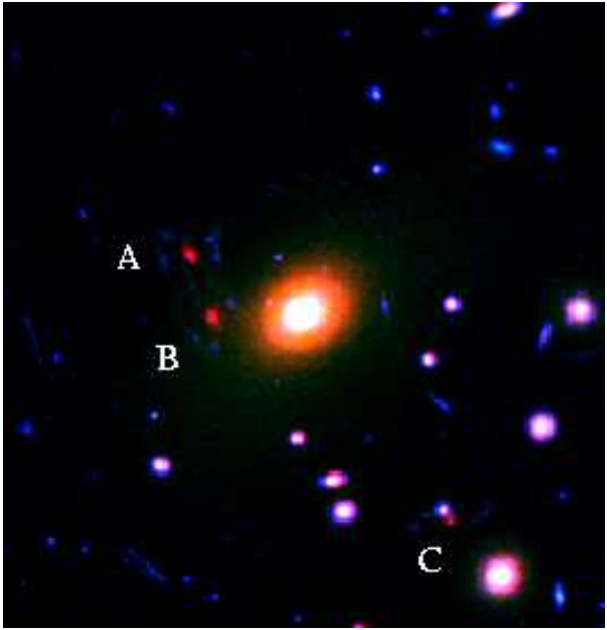,width=80mm}}
\caption{ True colour $RK$ view  ($50''\times50''$) of the core of the
galaxy cluster  A\,68, exploiting  the superlative resolution  of both
our  optical \emph{HST}  and  near-infrared UKIRT  imaging data.   The
bright elliptical  galaxy in  the centre of  the frame is  the central
galaxy  of the  cluster.   Three images  of  ERO\,J003707 are  clearly
visible and are marked A,B and  C.  Each image comprises a central red
``bulge'',  surrounded by numerous  fainter blue  knots of  current or
recent star formation.  North is up and East is left.  }
\end{figure}

The model  parameters were constrained  using the three images  of the
central red region  of ERO\,J003707, as the appearance  of these three
images  in the  $K$--band frame  allows us  to  unambiguously identify
these as images of the same underlying region of the galaxy.  We use a
$\chi^2$  estimator to  quantify  how  well our  lens  model fits  the
observational data (K96).  This estimator is minimised ($\chi^2\sim1$)
by varying the parameters of the model.

\section{Source-plane Reconstruction and Analysis}

%
%
\begin{figure}
\centerline{\psfig{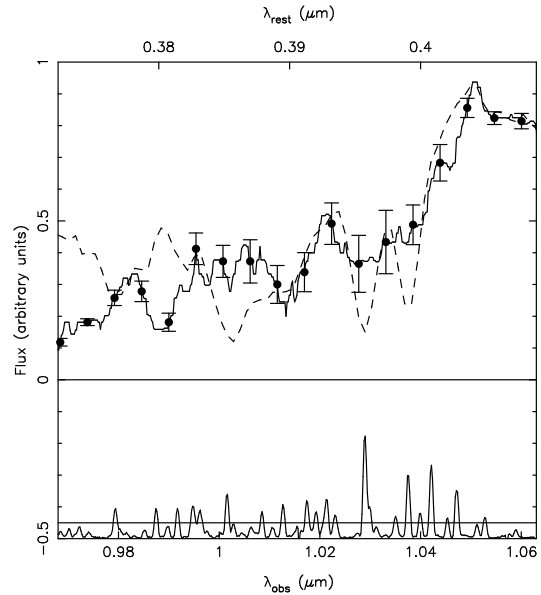}}
\caption{ The NIRSPEC $z$--band continuum  of the central red bulge of
ERO\,J003707,  produced by  median filtering  the  spectrum, rejecting
pixels dominated by strong night-sky emission i.e. those regions above
the horizontal line  in the lower spectrum, which  shows the night sky
spectrum (offset  and scaled  for clarity).  The  width of  the median
filter dominates the  uncertainty in the redshift quoted  in the text.
We   identify   the   discontinuity   in   the   spectral   shape   at
$\lambda=1.04\pm0.01\mu$m as the  $4000{\rm\AA}$ break.  The continuum
data are also plotted  as filled circles at $\Delta\lambda=50{\rm\AA}$
intervals, and we  estimate the uncertainties in these  data points by
bootstrap  resampling within  each median  filter window.   The dashed
line shows  a template spectrum of  a passive galaxy  from Mannucci et
al.\ (2001).  }
\end{figure}

We use  our lens model  to reconstruct the source-plane  properties of
ERO\,J003707.   We  threshold  the   $R$--  and  $K$--band  frames  at
1.5--$\sigma$ above the sky background and ray-trace each pixel to the
source   plane  at   $z=1.6$,  thus   creating  $R$--   and  $K$--band
source-plane  maps  of ERO\,J003707.   We  show,  as  an example,  the
source-plane reconstruction of ERO\,J003707$+$0909.5 (image ``A'' from
Fig.~1)  in Fig.~3b.   We  also simulate  blank-field observations  of
ERO\,J003707 (i.e.\ without a magnifying gravitational lens) by adding
the source reconstructions to noise maps from blank sky regions in the
original science frames (Figs.~3c~\&~3d).

%
%
\begin{figure*}
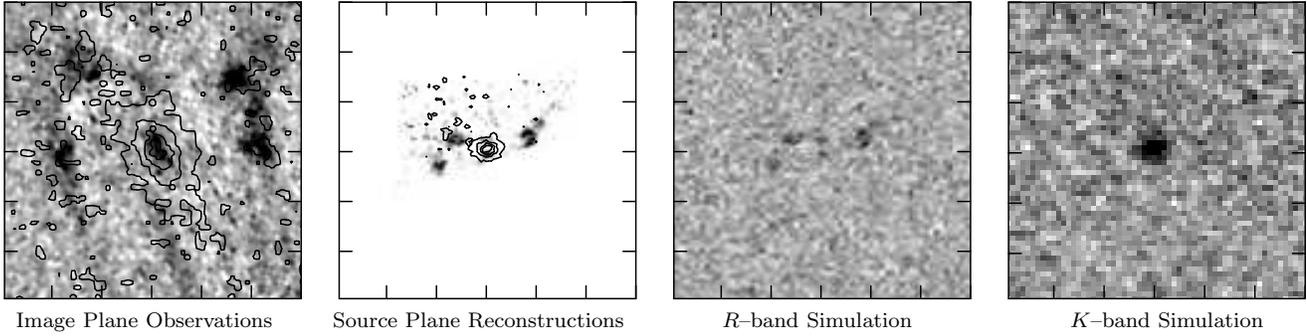


\centerline{
\psfig{file=figs/a68RKimg.ps,width=40mm,angle=-90}
\hspace{2.5mm}
\psfig{file=figs/a68RKsrc.ps,width=40mm,angle=-90}
\hspace{2.5mm}
\psfig{file=figs/a68Rsim.ps,width=40mm,angle=-90}
\hspace{2.5mm}
\psfig{file=figs/a68Ksim.ps,width=40mm,angle=-90}
}

\noindent\small{\hspace{-4mm}Image Plane Observations\hspace{8mm}Source 
Plane Reconstructions\hspace{13mm}$R$--band Simulation\hspace{20mm}
$K$--band Simulation}

\caption{ We illustrate the  morphology of ERO\,J003707 using the most
strongly amplified of  its three gravitational images --  image A from
Fig.~1.  (a)  The image-plane  morphology, showing the  observed F702W
morphology  (FWHM$\sim0.15''$)  as  the  grey-scale and  the  observed
$K$--band  morphology (FWHM$\sim0.4''$) as  contours.  The  light from
the central  galaxy of A\,68 had  been subtracted from  both the $R$--
and  $K$--band  data  in  this  panel.  (b)  The  reconstructed  F702W
(grey-scale)  and  $K$--band  (contours)  source-plane  morphology  of
ERO\,J003707;  both reconstructions are  displayed at  the pixel-scale
and  resolution  of  the  original  \emph{HST}  observations.   (c)  A
simulated   blank-field   7.5--ks   \emph{HST}/WFPC2  observation   of
ERO\,J003707.   (d)   A  simulated  blank-field   8.8--ks  UKIRT/UKIRT
(FWHM$\sim0.4''$) observation of ERO\,J003707.  Each panel is $6''$ on
a side and North is up and East in left.  }
\end{figure*}

We measure  the source-plane (i.e.\  ``un-lensed'') apparent $K$--band
magnitude  of  ERO\,J003707:  $K=19.8\pm0.1$.   This  translates  into
$M_K\sim-25.0$  (Fioc \& Rocca-Volmerange  1997), which  suggests that
the $K$--band  luminosity of ERO\,J003707  is comparable to  a present
day $L^{\star}$ galaxy (adopting $M_K^{\star}=-24.8$ from Cole et al.\
2001).   We  also  measure  the  $(R-K)$  colour,  using  an  aperture
equivalent to a $2''$--diameter in the source-plane.  We transform the
F702W  photometry to  Cousins $R$--band  and correct  for interstellar
extinction (see S02a for details), to give $(R-K)=5.4\pm0.1$.  This is
$\sim1$\,mag bluer than measured by  S02a because they did not correct
their  photometric aperture  for lens  amplification.   S02a therefore
measured  the colour  of  the central  red  bulge in  contrast to  our
photometry that also samples  the rest-frame ultraviolet emission from
the  disk.   We  also  compare  our photometry  with  spectral  energy
distributions  of   local  galaxies   using  {\sc  pegase}   (Fioc  \&
Rocca-Volmerange 1997), and find that  if a present-day Sa galaxy were
placed  at  $z=1.6$,  it  would  have  a  similar  $(R-K)$  colour  to
ERO\,J003707.

The  {\it  HST}  frame  reveals  a complex  and  irregular  rest-frame
ultraviolet  ($\sim2700{\rm \AA}$)  morphology, comprising  5--6 knots
indicating current  or recent star formation  (Fig.~3a).  In contrast,
the $K$--band  morphology consists of a  single centrally concentrated
component,  although there  is  some diffuse,  low surface  brightness
$K$--band  flux coincident  with the  knots of  rest-frame ultraviolet
emission  (Fig~3a).   Together with  our  spectroscopy  (\S2) and  the
robustness of  S02a's classification of  ERO\,J003707 as an  ERO after
correction for gravitational  lensing, this suggests that ERO\,J003707
is  a  disk-galaxy with  a  passively  evolving  central bulge  and  a
modestly star-forming  disk.  We quantify the  morphology by measuring
the source-plane  concentration ($C$) and asymmetry  ($A$) (Abraham et
al.\ 1996)  of the rest-frame ultraviolet and  $I$--band emission from
ERO\,J003707  using  our  $R$--  and  $K$--band  data.   We  estimate:
$C(2700{\rm\AA})=0.13\pm0.02$,           $C(8500{\rm\AA})=0.59\pm0.09$,
$A(2700{\rm\AA})=0.79\pm0.05$    and    $A(8500{\rm\AA})=0.56\pm0.04$.
These measurements  place ERO\,J003707 in the same  region of $C$--$A$
space  as  early-type  spiral-galaxies  in  the  local  Universe  when
observed at  ultraviolet and  optical wavelengths (Burgarella  et al.\
2001; Kuchinski et  al.\ 2001).  We also measure  the effective radius
of the central bulge ($r_{\rm e}$) and the exponential scale-length of
the disk ($r_{\rm d}$).  At $z=1.6$, the $4000{\rm\AA}$ break falls in
the $z$--band  (\S2) and so  our $R$-- and $K$--band  observations are
dominated by the stellar disk  and the central bulge respectively.  We
therefore  use our  $R$-- and  $K$--band  source-plane reconstructions
(Fig.~3b)   to   estimate    $r_{\rm   e}\sim0.6$\,kpc   and   $r_{\rm
d}\sim3.4$\,kpc.   We  compare these  measurements  with the  observed
$r_{\rm e}$--$r_{\rm  d}$ correlation for  early-type disk-galaxies in
the local  Universe (e.g.\  Khosroshahi et al.\  2000), and  find that
ERO\,J003707  is  consistent   with  this  correlation.   Finally,  we
estimate  that  the  rest-frame  $I$--band  bulge-to-total  luminosity
ratio, obtaining  ${\rm B/T}\sim0.80$ --  confirming that ERO\,J003707
is a bulge-dominated galaxy.

In  summary, our source-plane  photometric and  morphological analyses
reveal that ERO\,J003707  is similar to an $L^{\star}$  Sa galaxy from
the  local   Universe  that  is   observed  at  a   look-back-time  of
$\sim10^{10}$\,years.

\section{Diversity of Passive EROs}

The extremely  red colour  of ERO\,J003707 is  dominated by  a central
bulge of  old stars,  however ERO\,J003707 also  appears to  contain a
weakly star-forming  disk.  Contrary to the  popular assumption (\S1),
passive EROs therefore appear to comprise both elliptical galaxies and
early-type  spiral-galaxies at $z\gs1$  (Sa-EROs).  Such  Sa-EROs have
probably not  been identified in conventional  blank-field ERO surveys
(e.g.\ Daddi  et al.\ 2000) due  to the low surface  brightness of the
disk in these systems -- the disk is only detected at $\ls2$--$\sigma$
in our simulated blank-field observation of ERO\,J003707.

We attempt to constrain the  surface density of Sa-EROs.  Our analysis
of ERO\,J003707  reveals a colour gradient (\S4);  we therefore search
for this signature in the other 59 members of S02a's sample.  We first
quantify the expected gradient using our simulated $R$-- and $K$--band
blank-field observations of ERO\,J003707 to measure its seeing-matched
(FWHM$\sim0.4''$)   $1''$--  and   $2''$--diameter   aperture  $(R-K)$
colours.  Although  the $1''$--colour is  $0.6\pm0.3$\,mag redder than
the $2''$--colour, the two  colours are formally consistent within the
uncertainties.  We repeat this  measurement ten times, each time using
simulated  images  that   incorporate  different  noise  maps.   These
experiments  confirm the $\ls2$--$\sigma$  significance of  the colour
gradient in  our simulated observations.   We measure the  $1''$-- and
$2''$--diameter  aperture $(R-K)$  colours of  S02a's ERO  sample.  We
find that four  EROs display bluer colours at  larger radii consistent
with the simulations described above  and we note that S02a classified
all  four of  these galaxies  as having  an irregular  morphology.  We
therefore suggest  that $\sim10$ per  cent of EROs  with $(R-K)\ge5.3$
and $K\le21$ may  be Sa-EROs, which translates into  a surface density
of $\sim0.2\,{\rm arcmin}^{-2}$, based on S02a's counts.

\section{Evolution of ERO\,J003707}

We  speculate  on  the  evolution  of  ERO\,J003707.   The  rest-frame
$I$--band     luminosity      of     the     central      bulge     is
$\sim4.3\times10^{10}L_{\odot}$.   The   strong  $4000{\rm\AA}$  break
(\S2)  suggests that  the bulge  stars  in ERO\,J003707  formed a  few
billion years prior to the  epoch of observation (e.g.\ Dunlop et al.\
1996).  We estimate the age  of these stars using the ${\rm D}_{4000}$
index.  We  measure ${\rm D}_{4000}=2.4\pm0.5$,  which translates into
an  age of $\ge2.5$\,Gyr  (Poggianti \&  Barbaro 1997;  assuming solar
metallicity), implying a formation  redshift of $z_{\rm f}\gs3.7$ in a
$\Lambda$CDM  cosmology,  and  $z_{\rm  f}\gs6.5$ in  a  standard  CDM
cosmology (\S1).  This suggests that  the bulge was formed in a short,
but  intense starburst  at  $z_{\rm f}\gg2$.   This  event would  have
certainly  disrupted and  may have  destroyed any  existing  disk.  In
contrast  to  its  observed  morphology,  ERO\,J003707  was  therefore
probably a diskless galaxy at $z\sim3$.

Turning to  the disk component,  we estimate the  rest-frame $I$--band
disk luminosity  to be $\sim7.5\times10^{9}L_{\odot}$  and the current
star  formation  rate   (SFR)  to  be  SFR\,$\sim6{\rm  M}_{\odot}{\rm
yr}^{-1}$ (Kennicutt's 1998; assuming a Salpeter (1955) IMF integrated
over  0.1--100\,M$_{\odot}$ and  neglecting dust).   Ignoring mergers,
the future  evolution of ERO\,J003707  will depend on the  duration of
star formation in its disk.   If the gas becomes exhausted or expelled
from the galaxy, then the  star formation would cease and ERO\,J003707
would probably evolve into an E/S0  galaxy by the present day.  On the
other hand,  semi-analytic models of  galaxy formation (Baugh  et al.\
1996) predict  that bulges and  spheroids are  formed from  merging of
disk-galaxies and subsequently may accrete gas, thus re-growing a disk
component.   We   explore  this  ``transvestite   galaxy''  hypothesis
(Richard  Bower, private  communication) by  estimating  the timescale
($\tau_{\rm eq}$)  over which,  at the current  SFR, the  disk stellar
mass will become  comparable with the bulge stellar  mass.  We convert
the bulge luminosity  to a stellar mass (assuming  $M/L\sim1$; Bell \&
de  Jong  2001)  and divide  by  the  disk  SFR to  obtain  $\tau_{\rm
eq}\sim6\times10^{9}$\,yrs   which    is   roughly   equal    to   the
look-back-time  from the  present  day to  $z=1.6$.   This raises  the
intriguing possibility that the  progenitors of a fraction of luminous
spiral  galaxies in  the  local  Universe could  have  been EROs  when
observed at $z\gs1$.

\section{Conclusions}

We present  near-infrared spectroscopy and  optical \emph{HST} imaging
of      ERO\,J003707$+$0909.5,     the     brightest      of     three
gravitationally-lensed  images of  ERO\,J003707  in the  field of  the
massive  cluster A\,68  ($z=0.255$).   Our near-infrared  spectroscopy
reveals  a  break  at  $\lambda_{\rm  obs}=1.04\pm0.01\mu  m$  in  the
spectral energy distribution of this galaxy.  This feature arises from
the redshifted  $4000{\rm\AA}$ break  of a centrally  concentrated old
stellar population, which  places ERO\,J003707 at $z=1.60\pm0.03$.  We
constrain a  detailed model of  the cluster lens and  then reconstruct
the source-plane properties of ERO\,J003707.  Our main conclusions are
as follows:

\smallskip

\noindent (1)  The luminosity and  $(R-K)$ colour of  ERO\,J003707 are
similar to those of an $L^{\star}$  galaxy with a SFR comparable to an
Sa galaxy in the local Universe.

\smallskip

\noindent (2) Morphological parameters ($C$, $A$, $r_{\rm e}$, $r_{\rm
d}$,  $B/T$) based  on rest-frame  ultraviolet and  $I$--band emission
confirm that ERO\,J003707 is an early-type disk-galaxy.

\smallskip

\noindent (3)  The unique association of passive  EROs with elliptical
galaxies  is clearly too  simplistic.  We  estimate that  $\sim10$ per
cent  of EROs  with $(R-K)\ge5.3$  and $K\le21$  may be  Sa-EROs i.e.\
early-type disk-galaxies similar to ERO\,J003707.

\smallskip

\noindent (4) We estimate that  the bulge stars in ERO\,J003707 formed
$\ge2.5$\,Gyr  ago  and  speculate  that,  if  the  current  disk  SFR
($\sim6{\rm M}_{\odot}{\rm yr}^{-1}$) continues  due to the cooling of
gas onto  ERO\,J003707 in the manner predicted  by hierarchical galaxy
formation models (Baugh et  al.\ 1996), then ERO\,J003707 would evolve
into a luminous spiral galaxy by the present day.

\section*{Acknowledgments}

GPS thanks Richard Bower for his encouragement and Olga Kuhn for her 
assistance with the UKIRT/CGS4 observations.  
We are grateful to Carlton Baugh, Arjun Dey, Alastair Edge and Rob 
Ivison for helpful discussions and assistance.  
We acknowledge an anonymous referee for exceptionally prompt feedback.
We also thank Andy Adamson, John Davies and Tom Kerr for their support 
of the UKIRT observing programme.
GPS acknowledges a postgraduate studentship from PPARC. 
IRS acknowledges support from the Royal Society and a Philip Leverhulme 
Prize Fellowship.
JPK acknowledges support from CNRS.  
OC acknowledges support from the European Commission under
contract no.\ ER-BFM-BI-CT97-2471.
HE acknowledges support by NASA and STScI grants NAG 5--6336 and
GO 5--08249.
We also acknowledge support from the UK--French ALLIANCE collaboration 
programme \#00161XM.


\begin{thebibliography}{99}
\bibitem{} Abraham R.G., van den Bergh S., Glazebrook K., Ellis R.S., 
	Santiago B.X., Surma P., Griffiths R.E., 1996, ApJS, 107, 1{}{}
\bibitem{} Afonso J., Mobasher B., Chan B., Cram L., 2001, ApJ, 559, L101{}{}
\bibitem{} Baugh C.M., Cole S., Frenk C.S., 1996, MNRAS, 283, 1361{}{}
\bibitem{} Bell E.F.\ \& de Jong R.S., 2001, ApJ, 550, 212{}{}
\bibitem{} Bolzonella M., Miralles J.M., Pell\'o, 2000, A\&A, 363, 476{}{}
\bibitem{} Burgarella D., Buat V., Donas J., Milliard B., Chapelon, S., 
	2001, A\&A, 369, 421{}{}
\bibitem{} Cole S., Lacey C.G., Baugh C.M., Frenk C.S., 2000, MNRAS, 
	319, 168{}{}
\bibitem{} Cole S., Norberg P., Baugh C.M., Frenk C.S., \& the 2dFGRS team,
	2001, MNRAS, 326, 255{}{}
\bibitem{} Coleman G.D., Wu C.-C., Weedman D.W., 1980, ApJS, 43 393{}{}
\bibitem{} Daddi E., Cimatti A., Pozzetti L., Hoekstra H., Roettgering H., 
	Renzini A., Zamorani G., Manucci F., 2000, A\&A, 361, 535{}{}
\bibitem{} Dey A., Graham J., Ivison R., Smail I., Wright G., Liu M., 
	1999, ApJ, 519, 610{}{}
\bibitem{} Dunlop J., Peacock J., Spinrad H., Dey A., Jimenez R., Stern 
	D., Windhorst R., 1996, Nature, 381, 581{}{}
\bibitem{} Ebbels T.M.D., LeBorgne J.-F., Pell\'o R., Ellis R.S., Kneib 
	J.-P., Smail I., Sanahuja B.,  1996, MNRAS, 281, L75{}{}
\bibitem{} Eggen O.J., Lynden-Bell D., Sandage A. R., 1962, ApJ, 136, 748{}{}
\bibitem{} Ellis R.S., Santos M.R., Kneib J.-P., Kuijken K., 2001, ApJ, 
	560, L119{}{}
\bibitem{} Fioc M.\& Rocca-Vomerange B., 1997, A\&A, 326, 950{}{}
\bibitem{} Fontana A., Menci N., D'Odorico S., Giallongo E., Poli F., 
	Cristiani, S., Moorwood A., Saracco P., 1999, MNRAS, 310, L27{}{}
\bibitem{} Franx M., llingworth G.D.. Kelson D.D., van Dokkum P.G., 
	Tran K.-V., 1997, ApJ, 486, L75{}{}
\bibitem{} Hammer F.\ \& Rigaut F., 1989, A\&A, 226, 45{}{}
\bibitem{} Kauffmann G.\ \& Charlot S., 1998, MNRAS, 297, L23{}{}
\bibitem{} Kennicutt R.C., 1998, ARA\&A, 36, 189{}{}
\bibitem{} Khosroshahi H.G., Wadadekar Y., Kembhavi A., 2000, ApJ, 533, 
	162{}{}
\bibitem{} Kneib J.-P., Ellis R.S., Smail I., Couch W.J.,
	Sharples R.M., 1996, ApJ, 471, 643 (K96){}{}
\bibitem{} Kuchinski L.E., Madore B.F., Freedman W. L., Trewhella M., 
	2001, AJ, 122, 729{}{}
\bibitem{} Larson R.B., 1975, MNRAS, 173, 671{}{}
\bibitem{} McLean I.S., Becklin E.E., Bendiksen O., Brims G., Canfield J.,
	Figer D.F., Graham J.R., et al., 1998, SPIE, 3354, 566{}{}
\bibitem{} Mannucci F., Basile F., Cimatti A., Daddi E., Poggianti B.M., 
	Pozzetti L., Vanzi L., 2001, MNRAS, 326, 745{}{}
\bibitem{} Mountain C.M., Robertson D., Lee T.J., Wade R., 1990, 
	in Instrumentation in Astronomy, SPIE v1235, VII, 25{}{}
\bibitem{} Pell\'o R., Kneib J.-P., Le Borgne J.-F., Bézecourt J., 
	Ebbels T. M., Tijera I., Bruzual G., et al., 1999, A\&A, 346, 359{}{}
\bibitem{} Pierre M., Lidman C., Hunstead R., Alloin D., Casali M., 
        Cesarsky C., Chanial P., et al., 2001, A\&A, 372, L45{}{}
\bibitem{} Poggianti B.M. \& Barbaro G., 1997, A\&A, 325, 1025{}{}
\bibitem{} Salpeter E.E., 1955, ApJ, 121, 161{}{}
\bibitem{} Seitz S., Saglia R. P., Bender R., Hopp U., Belloni P., 
	Ziegler B., 1998, MNRAS, 298, 945{}{}
\bibitem{} Smith G.P., Kneib J-P., Ebeling H., Csozke O., Smail I., 2001a, 
	ApJ, 552, 493{}{}
\bibitem{} Smith G.P., Treu T., Ellis R., Smail I., Kneib J.-P., Frye B.L., 
	2001b, ApJ, 562, 635{}{}
\bibitem{} Smith G.P., Smail I., Kneib J.-P., Czoske O., Ebeling H., Edge 
	A.C., Pello R., et al., 2002a, MNRAS, 330, 1 (S02a){}{}
\bibitem{} Soifer B.T., Matthews K., Neugebauer G., Armus L., Cohen J.G., 
	Persson S.E., Smail I., 1999, AJ, 118, 2065{}{}
\bibitem{} Tinsley B.M.\ \& Gunn J.E., 1976, ApJ, 203, 52{}{}
\bibitem{} White S.D.M.\ \& Frenk C.S., 1991, ApJ, 379, 52{}{}
\end{thebibliography}
\end{document}